\documentstyle[12pt]{article}
\begin{document}

\vspace*{0.5cm}
\centerline{PACS no.s 04.60.+n, 03.65.-w, 11.10.-z}
\centerline{Preprint IBR-TH-99-S-01, Jan. 3, 1999}

\vspace*{1cm}
\begin{center}
{\bf PROBLEMATIC ASPECTS OF STRING THEORIES AND THEIR POSSIBLE
RESOLUTION}
\end{center}
\centerline{{\bf Ruggero Maria Santilli}}
\centerline{Institute for Basic Research}
\centerline{P.O.Box 1577, Palm Harbor, FL 34682, U.S.A.}
\centerline{ibr@gte.net; http://home1.gte.net/ibr}

\begin{abstract}
We identify new, rather serious,
physical and axiomatic inconsistencies of the current formulation of
string theories due to the lack of invariant units necessary for
measurements, lack of preservation in time of
Hermiticity-observability,  and other shortcomings. We propose three
novel reformulations of string theories for {\it matter} of
progressively increasing complexity via the novel iso-, geno- and
hyper-mathematics of hadronic mechanics, which  resolve the current
inconsistencies, while offering new intriguing possibilities, such as:
an axiomatically consistent and invariant formulation on curved
manifolds, the reduction of macroscopic irreversibility to the most
primitive level of vibrations of the universal substratum (ether), or
the
treatment of multi-valued biological structures. We then identify
three corresponding {\it classical}  formulations of string theories for
{\it antimatter} via the novel anti-isomorphic isodual mathematics. We
finally outline the intriguing features of the emerging new cosmologies
(including biological structures, as it should be for all
cosmologies), such as: universal invariance (rather than covariance)
under a symmetry isomorphic to the Poincar\'e group and its isodual;
equal distributions of matter and antimatter in the universe (as a limit
case); continuous creation; no need for the missing mass; significantly
reduced dimensions; possibility of experimental identification of
matter and antimatter in the universe; and identically null total
characteristics of  time, energy, linear and angular momentum, charge,
etc.
\end{abstract}

\vskip 0.50 cm

\noindent As it is well known, the origin of the physical consistency of
relativistic quantum mechanics (RQM) is its Lie structure, which we
express for subsequent needs with the following finite form,
infinitesimal version and conjugation,
$$
A(w) = U\times A(0)\times U^{\dagger} = e^{iX\times w}\times A(0)\times
e^{-iw\times X} = e_>^{iX>w} > A(0) < e_<^{-iw<X},
$$
$$
idA / dw = A\times X - X\times A = A < X - X > A,
$$
$$
e_>^{iX>w} = [ e_<^{- i w<X} ]^{\dagger},
\eqno (1)
$$
realized via {\it unitary transformations}
defined on a Hilbert space $\cal
H$ over  the field $C(c,+,\times)$ of complex numbers c with
conventional  sum + and associative) product $\times$.

In fact, the unitary structure implies the following well known basic
invariances:
$$
U\times U^{\dagger} =
U^{\dagger}\times U = I,
$$
$$
I\rightarrow U\times I\times U^{\dagger} = I' = I,
$$
$$
A\times B\rightarrow U\times (A\times B)\times U^{\dagger} =
(U\times A\times U^{\dagger})\times (U\times B\times U^{\dagger}) =
A'\times B',
$$
$$
H\times |\psi > = E\times |\psi>\rightarrow U\times H\times |\psi > =
(U\times H\times U^{\dagger})\times(U\times |\psi >)
= H'\times |{\psi}'> =
$$
$$
U\times E\times |\psi> = E'\times |{\psi}'>, E' = E.
\eqno (2)
$$

It then follows that {\it all theories with a unitary
structure defined on a Hilbert space over the field of complex numbers
possess numerically invariant units, products and
eigenvalues, thus being suitable to represent physical reality}.

By comparison, theories with a {\it nonunitary structure} have
serious flaws studied in details in Refs. [1], because invariances (2)
are turned  into the following noninvariances,
$$
U\times U^{\dagger} =
U^{\dagger}\times U \not = I,
$$
$$
I\rightarrow U\times I\times U^{\dagger} = I' \not = I,
$$
$$
A\times B\rightarrow U\times (A\times B)\times U^{\dagger} =
 (U\times A\times U^{\dagger})\times (U\times
U^{\dagger})^{-1}\times (U\times B\times U^{\dagger}) =
A'\times T\times B', T = (U\times U^{\dagger})^{-1},
$$
$$
H\times |\psi > = E\times |\psi>\rightarrow U\times H\times |\psi > =
(U\times H\times U^{\dagger})\times (U\times U^{\dagger})^{-1}\times (U
\times |\psi >)
=
$$
$$
H'\times T\times |{\psi}'> =
U\times E\times |\psi> = E'\times |{\psi}'>, E' \not = E,
\eqno (3)
$$
which imply rather serious {\it physical
inconsistencies}, such as: the lack of invariance of the basic units of
time, space, energy, etc, which is necessary for consistent
measurements; the lack of preservation in time of Hermiticity, which is
necessary to have physically acceptable observables; lack of uniqueness
and invariance in time of the physical predictions of the theory, and
other flaws (for details, see Refs. [1]).

Invariances (3) also have seemingly catastrophic axiomatic
inconsistencies at both classical and operator levels. Recall that all
axiomatic structures of physical theories (such as vector and metric
spaces, functional analysis, algebras and groups, etc.) are formulated
over a given field of numbers which, in turn, is crucially dependent on
the unit. The alteration in time of the basic unit then implies the loss
of the original field at subsequent times. This is due to the fact that
the noncanonical-nonunitary transform must be applied, for consistency,
to the {\it totality} of the original structure (see below), including
numbers,  and cannot be applied only to part of the original structure
to please personal preferences.  However, noncanonical-nonunitary
theories continue to be generally expressed over the original field. The
lack of invariance of the basic unit then implies the inapplicability of
the entire axiomatic structure of noncanonical-nonunitary
theories without any exception known to this author (for details see
Refs. [1g,3g]).

The above physical and axiomatic inconsistencies reach their climax for
all theories formulated on a {\it curved manifold} [1g,3g]. In fact,
the map from the minkowski metric $\eta = diag. (1, -1, -1, -1)$ =
constant to a Riemannian metric g(x) = function is transparently a {\it
noncanonical transform},
$\eta\rightarrow g(x) = U(x)\times \eta\times U^t(x), U(x)\times
U^t(x)\not = I$. Operator theories on curved manifolds must then be
necessarily nonunitary. As a result, the above physical and axiomatic
inconsistencies hold at both classical and operator levels (see
also [1g,3g] for brevity).

We therefore have the following
\vskip 0.50 cm

{\it THEOREM 1 [1]: All operator theories with a nonunitary structure
formulated  on a conventional Hilbert space
over the field of complex numbers,
including (but not limiting to) all
operator theories of gravity on a manifold with non-null curvature,
possess the following physical and axiomatic inconsistencies:

        I) lack of invariant units of space,
time, energy, etc.,
 with consequentially impossible
applications to real measurements as well as loss of the entire
axiomatic structure;

II) lack of preservation of the original Hermiticity in time, with
consequential
absence of physically acceptable observables;

III) general violation of probability and causality laws;

IV) lack of invariance of conventional and special
functions and transforms used in data elaborations;

V) lack of uniqueness and invariance of numerical predictions;

VI) General violation of the superposition principle
with consequential inapplicability to composite systems;

VII) General violation of Mackey imprimitivity theorem with
consequential  violation of Galilei's and Einstein's special
relativities.

All classical noncanonical theories formulated on conventional spaces
over conventional fields, including (but not limited to) all classical
theories of gravity formulated on a manifold with non-null curvature,
are
afflicted by corresponding physical and axiomatic inconsistencies which
prevent their consistent representation of physical reality.}
\vskip 0.50 cm

The above physical inconsistencies have been identified to occur for
numerous  theories, such as (see [1g] for details and literature):
1) Dissipative nuclear models with imaginary potentials;
2) Statistical models  with external collisions terms;
3) q-, k- and *-deformations;
4) Certain quantum groups (evidently those
with a nonunitary structure);
5) Weinberg's nonlinear theory;
6) All known theories of classical and quantum gravity on curved
manifolds;
7) All known supersymmetric theories;
8) All known Kac-Moody theories; and other theories.

In this note we point out apparently for the first time that, despite
an undeniable {\it mathematical} beauty, the {\it physical}
inconsistencies of Theorem 1 also  apply to current, classical and
operator formulations of {\it string theories}
  (see, e.g., Refs. [2] and vast literature quoted therein), because of
various reasons, such as:

A) The known nonunitary character of string theories formulated via the
Beta function according to Veneziano and Suzuki;

B) The more recent supersymmetric formulations of string
theories, because it implies the exiting from Lie's axioms (1) with
consequential  noninvariances (3);

C) Recent formulations of string theories on curved manifolds (see,
e.g., [2b]) because they  imply the additional, independent, rather
serious inconsistencies mentioned earlier.

The only way known to this author to resolve these inconsistencies
is that of reformulating string theories in such a way to
{\it regain the original invariances (2) in their totality}. In turn,
the only way known to this author to achieve such an objective is
the use of the new formalism of {\it hadronic mechanics} (see
Ref.s [3,4,5] and large literature quoted therein).

The invariant reformulations of {\it closed-isolated string systems}
requires the  isomathematics used in  the {\it isotopic branch of
hadronic mechanics}, and are here called {\it isotopic string theories}
(IST). Isomathematics is essentially based on lifting the
conventional n-dimensional unit
$I = diag. (1, 1, ..., 1)$ into a (nonsingular) $n\times
n$-dimensional quantity $\hat I$, called {\it isounit} (where the prefix
"iso-" means "axiom-preserving" character). In particular, $\hat I$
possesses an unrestricted functional dependence on time t,
coordinates r, momenta p, wavefunctions
$\psi$,  and any other needed variable. Jointly, the conventional
associative product
$A\times B$ among generic quantities A, B (such as numbers, vector
fields, operators, etc.) must be lifted into a form, called {\it
isoproduct}, which admits
$\hat I$ as the new right and left unit,
$$
I= diag.(1, 1, ..., 1)\rightarrow \hat I(t, r, p, \psi, ...) = U\times
U^{\dagger} =  1/\hat T
\not = I,
$$
$$ A\times B\rightarrow A\hat {\times}B = A\times
\hat T\times B,
$$
$$
\hat I\hat {\times} A = A\hat {\times} \hat I = A.
\eqno (4)
$$

The explicit construction of the IST then essentially requires:

a) The
identification of the background canonical-unitary theory, generally
consisting of RQM on a Minkowski space $M = M(x,\eta,F)$ with spacetime
coordinates x and metric $\eta = diag. (1, -1, -1, -1)$ on the reals $ F
= F(n,+,\times)$;

b) The identification of the {\it (noncanonical or)
nonunitary transform} of $\eta$ into the new metric g(x) of the
considered string theory,
$g(x) = U\times
\eta\times U^{\dagger}$;

c) The assumption of the isounit $\hat I = U\times U^{\dagger}\not = I$;

\noindent and the reconstruction of the totality of the
conventional formalism into such a form to admit $\hat I$ as the
correct left and right new unit, with no exception known to this
author.

This implies the lifting of [3f]: conventional numbers
and fields into the {\it
isonumbers} $\hat c$ and {\it isofields}
$\hat C(\hat c,\hat +, \hat {\times})$; conventional differential
calculus
into the isodifferential calculus;   conventional Hilbert spaces with
related states and inner product  into {\it
isohilbert spaces}
$\hat {\cal H}$ with {\it isostates} and {\it
isoinner product}; conventional eigenvalues equations
into {\it isoeigenvalue
equations}, etc., according to the rules
$$
\hat c = U\times c\times U^{\dagger} = c\times \hat I,
\hat c_1 \hat + \hat c_2 = (c_1 + c_2)\times \hat I, \hat c_1\hat
{\times}
\hat c_2 = (c_1\times c_2)\times \hat I,
$$
$$
\hat r^k = r^k\times \hat I, \hat d\hat r^k = \hat I^k_i \times d\hat
r^i,\hat {\partial}\hat {/}\hat {\partial}\hat r^k = \hat T_k^i\times
\partial/\partial r^i, \hat {\partial}\hat r^i\hat {/}\partial \hat r^j
= \hat {\delta}^i_k =
{\delta}^i_k\times \hat I,
$$
$$
|\hat {\phi}> = U\times |\phi>, U\times <\phi |
\times  |\psi>\times U^{\dagger} = < \hat {\phi}|\hat {\times}
|\hat {\psi}>\times \hat I,
$$
$$
U\times H\times |\phi> = \hat H\hat {\times} |\hat {\phi}> =
U\times E\times |\phi> = \hat E\hat {\times} |\hat {\phi}> = E\times
|\hat {\phi}>,
\hat H = U\times H\times U^{\dagger}.
\eqno (5)
$$

The transformation theory of the new string theory is then
strictly {\it nonunitary}. However, for consistency, all possible
nonunitary transforms must be rewritten as {\it isounitary
transforms} on $\hat {\cal H}$ over $\hat C$,  with consequential
regaining of all original  invariances (2) [3g], e.g.,
$$
V\times V^{\dagger} = \hat I \not = I, V = \hat V\times \hat T^{1/2},
V\times V^{\dagger} = \hat V\hat {\times} \hat V^{\hat {\dagger}} =
\hat V^{\hat {\dagger}}\hat {\times} \hat V = \hat I,
$$
$$
\hat I\rightarrow \hat V\times \hat I\times \hat V^{\hat {\dagger}} =
\hat I' = \hat I,
$$
$$
\hat A\hat {\times} \hat B\rightarrow \hat V\hat {\times}
(\hat A\hat {\times} \hat  B)\hat \times
\hat V^{\dagger} = V\times A\times V^{\dagger}\times V\times B\times
V^{\hat {\dagger}} = \hat A'\hat {\times}\hat B',
$$
$$
\hat H\hat {\times} |\hat {\psi} > =
\hat E\hat {\times} |\hat {\psi}>\rightarrow \hat V\hat {\times}
\hat H\hat \times |\hat {\psi} > =
\hat V\hat {\times}\hat  H\hat {\times}\hat V^{\hat {\dagger}}
\hat {\times}\hat V\hat {\times}
|\hat {\psi} > = \hat H'\hat {\times} |{\hat {\psi}}'> =
$$
$$
\hat V\hat {\times}\hat E\hat {\times} |\hat {\psi}> =
\hat E'\hat {\times}
|{\hat {\psi}}'>, \hat E' = \hat E,
\eqno (6)
$$

As one can see, the use of the isotopic formalism of hadronic mechanics
implies the full regaining of the {\it numerical invariance} of the
isounit, isoproduct and isoeigenvalues, thus regaining the
necessary conditions for physical applications. It is easy to prove that
{\it isohermiticity coincides with the conventional
Hermiticity}. As a result, all conventional observables of unitary
theories remain observables under their isotopic lifting. The
preservation of Hermiticity-observability in time is then ensured by
the above isoinvariances. Detailed studies conducted in Ref. [3g] then
establish the resolution of all inconsistencies of Theorem 1.

The primary reason for the consistency is the
full regaining of the Lie axioms. Again
under nonunitary transforms submitted to isotopic reformulation, we
have the rules
$$
U\times e^{X}\times U^{\dagger} = \hat e^{\hat X} = (e^{\hat X\times
\hat
T})\times \hat I = \hat I(e^{\hat T\times \hat X}),
$$
$$
\hat A(\hat w) =\hat U\hat {\times} \hat A(\hat 0)\hat {\times}
\hat U^{\hat {\dagger}} = \hat e^{i\hat X\hat {\times} w}\hat {\times}
\hat A(\hat 0)\hat {\times}\hat e^{-i\hat w\hat {\times}\hat X} =
e^{(\hat X\times w)\times \hat T}\times \hat A(\hat 0)\times
e^{-i\hat T\times (\hat w\times \hat X)},
$$
$$
i\hat d\hat A \hat {/}\hat  d\hat w = \hat A\hat {\times}\hat  X - \hat
X\hat {\times}\hat A = \hat A\times \hat T\times\hat  X - \hat X \times
\hat T\times \hat  A = [\hat A\hat {,}\hat  X],
$$
$$
\hat e^{i\hat X\hat {\times}\hat w} = [ \hat e^{- i \hat w\hat
{\times}\hat X} ]^{\hat {\dagger}}.
\eqno (7)
$$

As one can see, the regaining of Lie's theory is so strong that the
conventional and isotopic theories coincide at the abstract,
realization-free level. In fact, the Lie-Santilli isotopic theory
[3,4,5] can be formulated by essentially "putting the hat" to the {\it
totality} of symbols and operations of the conventional formulation of
Lie's theory or,  equivalently, by keeping the conventional
formulation and subjecting {\it all} conventional symbols to the more
general isotopic interpretation.

It should also be recalled that, since I and $\hat I$ are topologically
equivalent, the isotopic images of all Lie
groups are locally isomorphic to the original groups. This implies the
{\it preservation of the exact validity for nonunitary string theories
of  the fundamental spacetime symmetries, such as the Poincar\'e
symmetry, Einstein's special relativity and
well as relativistic quantum mechanics} (see Refs. [3h,3j] for details).

The above reformulation implies a new representation of
classical and operator gravity via an {\it isoflat geometry} (i.e.,
a geoometry flat on isospaces over isofields),  introduced by this
author
under the name of {\it isominkowskian geometry} [3j], which is defined
on
isospaces
$\hat M =
\hat M(\hat x,\hat {\eta}\hat F)$ with isocoordinates $\hat x = x\times
\hat I$ and isometric $\hat {\eta} =
\hat {\eta}(x, v, a, \psi, ...)$ on the isoreals $\hat F = \hat F(\hat
n,\hat +,\hat {\times})$.

The
new classical and operator formulation of gravity was first proposed in
Ref. [3i] (see memoir [3j] for a more recent treatment), in which
gravity
is merely embedded in the {\it unit}  of conventional, classical and
operator Minkowskian theories via the factorization of all possible
Riemannian metrics
$g(x) = \hat T_{grav}\times \eta$ and the assumption of the
gravitational isounit $\hat I_{grav} = \hat T_{grav}^{-1}$. Since
curvature is evidently contained in the term $\hat T_{grav}(x)$, the
assumption for fundamental unit of the {\it inverse},
$\hat I_{grav} = \hat T_{grav}^{-1}$, evidently eliminates curvature at
the abstract level. Moreover, $\hat T$ is necessarily positive-definite
(from the local Minkowskian character of Riemann). As a result, gravity
is formulated for the first time under a universal {\it invariance}
(rather than "covariance") isomorphicc to the Poinncar\'e symmetry [3h].

Also, the isominkowskian geometry is a symbiotic unification of the
Minkowskian and Rienmannian geometries because, on one side, it is
locally isomorphic to the Minkowskian geometry while, on the otehr side,
it preserves all the machinery of
Riemann (such as covariant derivatives, Christoffel's symbols, etc.),
although formulated via the isodifferential calculus (because
the isometric depends on x).

One obtains a new {\it classical iso-gravity} (CIG)
which preserves the conventional Einstein-Hilbert (and other) field
equations. Yet, the abandonment of the notion of curvature (which is
necessary to resolve the inconsistencies of Theorem 1) permite the
formulation of gravity as an {\it isocanonical theory} under the
universal isopoincar\'e symmetry, thus resolving the inconsistencies of
Theorem 1. Similarly, one obtains a new {\it operator iso-gravity}
(OIG)  which coincides at the abstract level with RQM, including
the operator version of the universal isopoincar\'e invariance,
thus resolving the inconsistencies of Theorem
1 at the operator level too.

The reconstruction of gravity on an isoflat space has permitted the
achievement of an axiomatically consistent grand unification of
electroweak and gravitational interactions first proposed in Ref.
[3k] (see [3l] for more details) under the name of {\it iso
grand-unification} (IGU), in which gravity is embedded in the {\it unit}
of $U(2)\times U(1)$. This grand unification is evidently available to
string theories in isotopic reformulation.

The explicit formulation of IST is elementary. Consider, for instance,
Ref. [2b]. Its basic assumption is metric (1.3), p. 50, in a curved
space $g = (1, -a^2\times {\delta}^i_j)$. It is evident that such a
metric is a noncanonical image of the conventional Minkowski metric
$g = U\times \eta\times U^t, U = diag. (1, a\times {\delta}^i_j)$. The
isotopic reformulation of such a theory then requires the use of the
isounit $\hat I = (1, a^2\times {\delta}^i_j)$ and the reconstruction
of the totality of the formalism of Ref. [2b] with respect to $\hat I$,
including numbers, fields, spaces, algebras, functional analysis, etc.
Invariance and the resolution of the   inconsistencies of Theorem 1 then
follow.

Note that the construction
implies a mere {\it reformulation} of conventional string theories
without altering the results.

As an incidental note, it should
be mentioned that {\it all papers on the isotopic branch of hadronic
mechanics prior to the appearance of the isodifferential calculus [3f]
in 1996 (beginning with the
articles by this author) generally have no physical applications}.
This is due to  the
lack of invariance of the basic dynamical equations because
they are formulated via the conventional differential calculus, even
though the rest of the theories is formulated on isospaces over
isofields [3g].

For the case of {\it open-irreversible string theories}, that is,
strings interacting with systems considered as external, we need the
more general {\it genotopic branch of hadronic mechanics} with a
Lie-admissible structure first proposed by this author in his Ph. D.
studies, Ref. [3b] of 1967, as a (p, q)-parametric deformation of
quantum
mechanics, $(A, B) = p\times A\times B - q\times B\times A$, and in Ref.
[3c] of 1978 as a (P, Q)-operator deformation, $(A\hat {,} B) =
A\times P\times B - B\times Q\times A$, where the prefix "geno" now
indicates an "axiom-inducing" character.

Lie-admissible theories reached
mathematical maturity only recently in memoir [3f] of 1996 and
invariance of physical formulations in the subsequent Ref. [3n] thanks
to
the advent of the {\it genodifferential calculus} of the preceding
memoir
[3f]. Therefore, {\it all papers on Lie-admissible theories prior to
Refs.
[3f,3n] (beginning with the papers by this author) generally have no
physical applications} because of the lack of invariance of the basic
dynamical equations due to their formulation via the conventional
differential calculus.

The construction of open-irreversible {\it genotopic
string theories} (GTS) requires {\it two different
nonunitary systems and related transforms}, one for the forward
direction of time
$>$ and one for the backward direction $<$.
Consequently, GST need {\it two} generalized units called {\it
genounits}, {\it two } products called {\it genoproducts} and
corresponding dual formalism, again, one per each direction of time,
along the following main lines
$$
V\times V^{\dagger}\not = I, W\times W^{\dagger} \not = I,
$$
$$
V\times W^{\dagger} = \hat I^> = 1/\hat S, \hat A>\hat B = \hat A\times
\hat S\times \hat B,
 \hat I^{>}>\hat A = \hat A>\hat I^> = \hat A,
$$
$$
W\times V^{\dagger} = ^<\hat I = 1/\hat R, \hat A<\hat B = \hat A\times
\hat R\times \hat B, ^<\hat I < \hat A = \hat A<^<\hat I = \hat A,
$$
$$
\hat A = \hat A^{\dagger}, \hat B = \hat B^{\dagger}, \hat R = \hat
{S}^{\dagger}.
\eqno (8)
$$

The above elements must then be completed,
for necessary reasons of consistency, with the forward and backward
genofields, genospaces, genodifferential calculus, genogeometries,
etc. [3f,3h,3r].

The procedure yields the following Lie-admissible realization of
Lie's axioms (1) at a fixed value of the parameter
w (thus without its ordering and by omitting the ordering in the
individual generators for simplicity of notation) [3c,3d]
$$
\hat A(\hat w) =  e_>^{i\hat X>\hat w} > \hat A(\hat 0) < e_<^{-i\hat
w<\hat X} = [e^{(i\hat X>\hat w)\times \hat S}\times \hat
{I}^>]\times \hat S\times \hat A(\hat 0)\times\hat R\times [^<\hat
{I}\times e_<^{-i\hat R\times (\hat w<\hat X}],
$$
$$
i \hat d\hat A\hat {/}\hat {d}\hat w = (\hat A, \hat X) = \hat A < \hat
X - \hat X > \hat A  = \hat A\times
\hat R\times \hat X - \hat X\times {\hat S}\times \hat A,
$$
$$
\hat X = \hat X^{\dagger}, \hat R = \hat {S}^{\dagger}.
\eqno (9)
$$

It should be stressed that structures (9) merely provide a {\it broader
realization} of the original Lie axioms (1), which is the basic theme
of hadronic mechanics [3]. In fact, the original Lie axioms (1) have a
{\it bimodular associative structure}, with a modular-associative
action to the right and a separate one to the left. These actions do not
need to be given
by the simplest conceivable realization of current use, because the
isotopic realization is equally admissible. The lack of necessary
identity of the two modular-isotopic actions then yields genotopic
structures (9), provided that conjugation (1c) is verified.

In Ref. [3a] of 1956 this author (then in high school) showed how the
"ethereal wind" used at the time to dismiss the existence of a
universal substratum (or ether) had no solid physical foundations
because said universal substratum is needed not only to propagate
electromagnetic waves, but also for the very existence of elementary
particles (such as the electrons), which are
"oscillations-vibrations" of the same medium. The author was unaware
at the time that, over twenty years earlier, Schroedinger had proved
that the variable "x" in Dirac's equations for the {\it free}  electron
describes precisely an oscillation which can only be that of the
universal; substratum. The transversal
character of electromagnetic waves demands that the universal
substratum be a rigid medium. In short, the view presented in Ref. [3a]
is that {\it space is completely full while
matter is completely empty.} When
matter is moved we merely transfer the oscillations of space from one
region to another, thus without any possible "ethereal wind".

The genotopic reformulation of string theories permits quantitative
studies along the view of Ref. [3a]. In fact,
{\it GST permit an axiomatically consistent and invariant reduction of
our macroscopic irreversible physical reality to
the most elementary entities in the universe, the vibrations of the
universal substratum under open-nonconservative conditions}.

As one can see, Lie-admissible structures (9) are {\it structurally
irreversible} in the sense that they are {\it irreversible}  for all
possible conventional,  reversible  Hamiltonians. This is precisely what
needed for a serious study of irreversibility because
all action-at-a-distance interactions are well known to be reversible,
while physical reality is irreversible.
Irreversibility should therefore be represented with anything {\it
except the Hamiltonian}. This is along the historical
teaching by Lagrange and Hamilton who represented irreversibility via
{\it the external terms} in their celebrated equations, which terms have
been "truncated" in the literature of this century. The use of two
different generalized units for the representation of irreversibility
appears to be preferable over other attempts, evidently because it
assures invariance.

Interested readers can then find in Refs. [3f,3n] the invariant
genotopic
formulation of: Newton's equations with contact-nonhamiltonian forces;
the true Hamilton equations (those with external terms); quantization;
and quantum mechanics. When applied to string theories, these
formulations then permit the indicated reduction of our classical
irreversibility to the most elementary possible entities,
open-nonconservative vibrations of space.

Physics is a science that will never admit "final theories". By no
means GST's are the most general ones. For completeness we
mention the existence of the {\it hyperstructural branch of hadronic
mechanics} [3f,3g] whose main characteristic is that of being {\it
multi-valued} with {\it hyperunits and hyperproducts},
$$
\hat I^> = \{\hat {I}_1^>, \hat {I}_2^>, \hat {I}_3^>, ...\} = 1/\hat
S,
$$
$$
A>B = \{ A\times \hat S_1\times B, A\times \hat S_2\times B,
A\times \hat S_3\times B, ...\},
 \hat I^{>}>A = A>\hat I^> = A\times I,
$$
$$
^<\hat I = \{^<\hat {I}_1, ^<\hat {I}_2, ^<\hat {I}_3, ...\} = 1/\hat S,
$$
$$
A<B = \{A\times \hat R_1\times B, A\times hat R_2\times B,
A\times \hat R_3\times B, ...\}
^<\hat I < A = A<^<\hat I = A\times I,
$$
$$
A = A^{\dagger}, B = B^{\dagger}, \hat R = \hat {S}^{\dagger}.
\eqno (10)
$$

All aspects of the dual Lie-admissible
formalism admit a unique, and significant extension to
the above hyperstructures, including {\it hypernumbers and fields,
hyperspaces, hyperdifferential calculus, etc.}

The construction of invariant {\it hyper-string theories} (HST) is then
elementary and will be left to the interested reader. Their most salient
(and intriguing) feature is that of {\it permitting the existence of
a multi-valued universe beginning at the most elementary level of
nature, and  in a way compatible with our three-dimensional sensory
perception.}

A suggestive illustration is given by {\it sea shells} [3r]. All their
possible {\it shapes} can indeed be fully represented in the
conventional Euclidean  space corresponding to our
three Eustachian tubes. However, their {\it evolution in time}
cannot be described via the Euclidean axioms, trivially, because sea
shells are open-nonconservative-irreversible, while the Euclidean axioms
are strictly closed-conservative-reversible. Computer visualizations
have shown that the imposition of the latter to the time evolution of
the former implies that sea shells first grow in a deformed way and
then they crack. Studies have shown that the
quantitative representation of the growth in time of sea shells requires
at least a {\it six dimensional space}, i.e., the doubling of each
reference axis.

However, we can directly observe sea shells in our hands
as being three-dimensional.
The only reconciliation of these seemingly
dissonant occurrences known to this author is that via
hyperformulations [3f]. In fact, the latter are multi-dimensional
precisely as needed for the representation of the growth of sea shells.
Yet the axioms remain conventional,
thus achieving compatibility with our sensory perception.

Specifically, for the case of sea shells we have
two-valued hypereuclidean formulations which are fully
compatible with our sensory perception because {\it
they are not six-dimensional}, and remain  instead fully
three-dimensional. After all, there is a dramatic topological difference
between a conventionally {\it six-dimensional} space and our {\it
two-valued three-dimensional} space.

The above example indicates the possibility of reducing
open-nonconservative-irreversible {\it biological} structures to the
ultimate vibrations of the universal substratum, thus extending
irreversibility from sole physical systems to include biology.

Note that any beliefs in treating the above
open-nonconservative-irreversible systems via conventional quantum
mechanics implies exiting science, evidently because of the strictly
closed-conservative-reversible character of the theory. This is the
main motivation for the construction of hadronic mechanics [3].

 By no means HST represent the ultimate and most general formulation of
string theory. In fact, despite their remarkable generality,
hyperformulations (including conventional, isotopic and, genotopic
particularizations)  cannot consistently
represent {\it antimatter} at the {\it classical} level. This is due to
the sole existence of {\it one} quantization channel, as a consequence
of
which {\it the operator images of classical iso, geno and
hyper-structures cannot yield charge conjugated antiparticles, but only
particles with the wrong sign of the charge}.

The above occurrence is only a symptom of what can be safely claimed to
be the biggest unbalance of theoretical physics of this century:
the treatment of matter at all possible levels, from Newton to second
quantization, while antimatter is solely treated at the level of {\it
second} quantization.

After a laborious search, the only {\it classical} representation of
{\it
antimatter} this author could identify is that characterized by the
following map, called {\it isoduality}, here expressed for a generic
quantity A, as well as for the underlying spaces and fields [3o],
$$
A(x, v, \psi, ...)\rightarrow A^d =
-A^{\dagger}(-x^{\dagger}, -v^{\dagger}, -\psi^{\dagger}, ...)
\eqno (11)
$$
which characterizes the {\it isodual branches of hadronic mechanics}
[3g].

The above map is mathematically nontrivial, e.g., because it implies the
first
known {\it numbers with negative units and norm}
[3e].
Physically, the map is also nontrivial because it implies an isodual
image of our
universe which coexists with our own, yet it is distinct from
it. Universes interconnected by isoduality are then {\it
anti-isomorphic}
to each others, as it is the case for the charge conjugation. In fact,
isoduality is equivalent to charge conjugation at the level of second
quantization [3o]. In particular,  {\it all physical quantities (and not
only the charge) interconnected by isoduality have opposite signs,
although referred to units also with opposite signs}. For instance, time
for an isodual antiparticle is {\it  negative,} although referred to a
{\it negative unit of time} -1 sec, thus being as causal as our
conventional positive time referred to the conventional positive unit +1
sec. The kronecker product of a universe and its isodual is called {\it
isoselfdual} in the sense of coinciding with its isodual image (see [3o]
for details).

Isotopic, genotopic and hyper-string theories therefore admit isodual
images whose explicit construction is left to the interested reader for
brevity. Their significance is not only restricted to a {\it classical}
reduction of {\it antimatter} to vibrations at the ultimate possible
level of the universe, but also that
of permitting a novel cosmology, here called {\it isoselfdual
hyperstring cosmology} (IHSC) along the lines of Ref. [3m] with rather
intriguing characteristics, such as:
1) definition of cosmology inclusive
of biological structures (as it should be under the meaning of the term
"cosmos");
2) same amount of matter and antimatter in the universe (as a
limit conditions under Lie axiom (1c));
3) open-irreversible structure of
the universe with continuous creation (except for the particular
isotopic
case);
4) universal isopoincar\'e {\it invariance} inclusive of gravitation
in isominkowskian reformulation [3l];
5) lack of need of the "missing
mass",  because the average speed of light c of galaxies and quasars in
the law $E = m\times c^2$ is {\it bigger} the speed of light in vacuum
$c_o$ whn including all interior grav itational problems [3m];
6) considerable reduction of the currently believed size of the
universe,
because light exits galaxies and quasars already redshifted due to the
decrease of its speed within the huge and hyperdense astrophysical
chromospheres;
7) possibility of future experimental study whether a
far away galaxy or quasar is made up of matter or of antimatter due to
the prediction that the photon emitted by antimatter, the {\it isodual
photon}, is repelled by gravity and has new parity properties [3o];
8) identically null total characteristics of time, mass, energy, etc.;
9) consequential identical topological features of the universe
befoire and after creation; and other intriguing features.

\end{document}